\documentclass[runningheads]{llncs}

\usepackage{url}
\usepackage{color}
\usepackage{paralist}
\usepackage{fancyvrb}
\usepackage{epsfig}
\usepackage{cite}
\usepackage{amsmath}
\usepackage{booktabs}

\newcommand{\q}[1]{\lq\lq{}{}#1\rq\rq{}{}}
\newcommand{\service}[0]{{\tt\scshape service}}
\newcommand{\argmin}{\operatornamewithlimits{argmin}}

\begin{document}

\title{Heuristics-based Query Reordering for Federated Queries in SPARQL 1.1 and SPARQL-LD}
\titlerunning{Heuristics-based Query Reordering for Federated Queries in SPARQL}

\author{
	Thanos Yannakis\inst{1} \and 
	Pavlos Fafalios\inst{2} \and 
	Yannis Tzitzikas\inst{1} }
\institute{
	Computer Science Department, University of Crete, and FORTH-ICS, Greece\\
	\email{\{yannakis, tzitzik\}@ics.forth.gr}
	\and
	L3S Research Center, Leibniz University of Hannover, Germany\\
	\email{fafalios@L3S.de}
}

\maketitle

\begin{abstract}
The federated query extension of SPARQL 1.1 allows executing queries distributed over different SPARQL endpoints.
SPARQL-LD is a recent extension of SPARQL 1.1
which enables to directly query any HTTP web source containing RDF data,
like web pages embedded with RDFa, JSON-LD or Microformats,
without requiring the declaration of named graphs.
This makes possible to query a large number of data sources
(including SPARQL endpoints, online resources, or even Web APIs returning RDF data)
through a single one concise query.
However, not optimal formulation of SPARQL 1.1 and SPARQL-LD queries
can lead to a large number of calls to remote resources
which in turn can lead to extremely high query execution times.
In this paper, we address this problem and propose a set of query reordering methods
which make use of heuristics to reorder a set of \service\ graph patterns based on
their restrictiveness, without requiring the gathering and use of statistics
from the remote sources.
Such a query optimization approach is widely applicable since it can be
exploited on top of existing SPARQL 1.1 and SPARQL-LD implementations.
Evaluation results show that query reordering can highly decrease the query-execution time,
while a method that considers the number and type of unbound variables and joins achieves the optimal query plan in 88\% of the cases.

\end{abstract}

\section{Introduction}

A constantly increasing number of data providers
publish their data on the Web following the Linked Data principles and
adopting standard RDF formats.
According to the Web Data Commons project \cite{muhleisen2012web},
38\% of the HTML pages in the Common Crawl\footnote{\url{http://commoncrawl.org/}} of October 2016
contains structured data in the form of RDFa, JSON-LD, Microdata, or Microformats\footnote{\url{http://webdatacommons.org/structureddata/2016-10/stats/stats.html}}.
This data comes from millions of different pay-level-domains,
meaning that the majority of Linked Data is nowadays available through a large number of different data sources.
The question is: how can we {\em efficiently} query this large, distributed, and constantly increasing body of knowledge?

SPARQL \cite{sparqlQueryLang} is the {\em de facto} query language
for retrieving and manipulating RDF data.
The SPARQL 1.1 Federated Query recommendation of W3C
allows executing queries distributed over different SPARQL endpoints \cite{sparqlFedQuery}.
SPARQL-LD \cite{fafalios2015sparql,fafaliosquerying} is an extension (generalization) of
SPARQL 1.1 Federated Query which extends the applicability of
the \service\ operator to enable querying any HTTP web source containing RDF data,
like online RDF files (RDF/XML, Turtle, N3) or web pages embedded with RDFa, JSON-LD, or Microformats.
Another important characteristic of SPARQL-LD is that it does not require the named graphs to have been declared,
thus one can even fetch and query a dataset returned by a portion of the query, i.e., whose URI is derived at query execution time.
Thereby, by writing a single concise query, one can query hundreds or thousands of data sources,
including SPARQL endpoints, online resources, or even Web APIs returning RDF data \cite{fafaliosquerying}.

However, not optimal query writing in both SPARQL 1.1 and SPARQL-LD
can lead to a very large number of \service\ calls to remote
resources, which in turn can lead to an extremely high query
execution time. Thus, there arises the need for an effective query
optimization method than can find a near-optimal query execution
plan. In addition, given the dynamic nature of Linked Data and the
capability offered by SPARQL-LD to query any remote HTTP resource
containing RDF data, we need a widely-applicable method that does
not require the use of statistics or metadata from the remote
sources and that can operate on top of existing SPARQL 1.1 and
SPARQL-LD implementations.

To this end, in this paper we propose and evaluate a set of query
reordering methods for SPARQL 1.1 and SPARQL-LD.
We focus on fully heuristics-based
methods that reorder a query's \service\ graph patterns based on their restrictiveness (selectivity),
without requiring the gathering
and use of statistics from the remote sources. The objective is to
decrease the number of intermediate results and thus the number of
calls to remote resources. We also propose the use of a greedy algorithm
for computing a near-optimal query execution plan for cases of large number of \service\ patterns.

In a nutshell, in this paper we make the following contributions:
\begin{compactitem}
\item   We propose a set of heuristics-based query reordering methods for SPARQL 1.1 and SPARQL-LD, which can also exploit a greedy algorithm for choosing a near-optimal query execution plan. The query optimizer is publicly available as open source.\footnote{\label{footnote:sourceCode}\url{https://github.com/TYannakis/SPARQL-LD-Query-Optimizer}}
\item   We report the results of an experimental evaluation which show that a method that considers the number and type of unbound variables and the number and type of joins achieves the optimal query plan in 88\% of the examined queries, while the greedy algorithm has an accuracy of 94\% in finding the reordering with the lowest cost.
\end{compactitem}
\vspace{1mm}

The rest of this paper is organized as follows:
Section \ref{sec:rw} presents the required background and related works.
Section \ref{sec:optim} describes the proposed query reordering methods.
Section \ref{sec:eval} reports experimental results.
Finally, Section \ref{sec:concl} concludes the paper and discusses interesting directions for future work.

\section{Background and Related Literature}
\label{sec:rw}

\subsection{SPARQL-LD}
\label{sec:sparqlld}

The \service\ operator of  SPARQL 1.1 (\service\ $a$ $P$) is defined
as a graph pattern $P$ evaluated in the SPARQL endpoint
specified by the URI $a$,
while (\service\ $?X$ $P$)
is defined by assigning to the variable $?X$
all the URIs (of endpoints) coming from partial results,
i.e. that get bound after executing an initial query fragment \cite{buil2013federating}.
The idea behind SPARQL-LD is to enable the evaluation of a graph pattern $P$ not absolutely to a SPARQL endpoint $a$,
but generally to an RDF graph $G_r$ specified by a Web Resource $r$.
Thus, now a URI given to the \service\ operator
can also be the dereferenceable URI of a resource,
the Web page of an entity (e.g., of a person),
an ontology (OWL), Turtle or N3 file, or even the URL of a service that
dynamically creates and returns RDF data.
In case the URI is not the address of a SPARQL endpoint,
the RDF data that may exist in the resource are fetched at real-time
and queried for the graph pattern $P$.
Currently, SPARQL-LD supports a variety of standard formats,
including RDF/XML, N-Triples, N3/Turtle, RDFa, JSON-LD, Microdata, Microformats \cite{fafalios2015sparql,fafaliosquerying}.

SPARQL-LD is a generalization of SPARQL 1.1 in the sense
that every query that can be answered by SPARQL 1.1 can be also
answered by SPARQL-LD.
Specifically, if the URI given to the \service\ operator corresponds to a SPARQL endpoint,
then it works exactly as the original SPARQL 1.1
(the remote endpoint evaluates the query and returns the result).
Otherwise, instead of returning an error (and no bindings),
it tries to fetch and query the triples that may exist in the given resource.
SPARQL-LD has been implemented using Apache Jena \cite{apachejena},
an open source Java framework for building Semantic Web applications.
The implementation is available as open source\footnote{\url{https://github.com/fafalios/sparql-ld}}.

Listing \ref{fig:introExample} shows a query
that can be answered by SPARQL-LD.
The query returns all co-authors of Pavlos Fafalios together with
the number of their publications and the number of distinct conferences in which they have a publication.
The query first accesses the RDFa-embedded web page of Pavlos Fafalios to collect his co-authors,
then queries a SPARQL endpoint over DBLP to retrieve the conferences, and finally
accesses the URI of all co-authors to gather their publications.
Notice that the co-author URIs derive at query-execution time.
In the same query, one could further integrate
data from any other web resource, or from a web API
which can return results in a standard RDF format.

The query in Listing \ref{fig:introExample} is answered within a few seconds. However, if we change the
order of the first two \service\ patterns, then its execution time is dramatically increased to many minutes.
To cope with this problem, in this paper we propose methods to reorder the query's \service\ patterns
and thus improve the query execution time in case of non optimal query formulation.

\renewcommand{\figurename}{Listing}
\setcounter{figure}{0}

\begin{figure}[th]
\vspace{-2mm} \centering \scriptsize
\begin{Verbatim}[frame=lines,numbers=left,numbersep=1pt]
 SELECT DISTINCT ?authorURI (count(distinct ?paper) AS ?numOfPapers)
                            (count(distinct ?series) AS ?numOfDiffConfs) WHERE {
   SERVICE <http://l3s.de/~fafalios/> { ?p <http://purl.org/dc/terms/creator> ?authorURI }
   SERVICE <http://dblp.l3s.de/d2r/sparql> {
      ?p2 <http://purl.org/dc/elements/1.1/creator> ?authorURI .
      ?p2 <http://swrc.ontoware.org/ontology#series> ?series }
   SERVICE ?authorURI { ?paper <http://purl.org/dc/elements/1.1/creator> ?authorURI }
 } GROUP BY ?authorURI ORDER BY DESC(?numOfPapers)
\end{Verbatim}
\vspace{-5mm}
\caption{Example SPARQL query that can be answered by SPARQL-LD.}
\label{fig:introExample}
\vspace{-2mm}
\end{figure}

\subsection{Related Works}
\label{sec:sparqlld}

\subsection*{SPARQL Endpoint Federation}
The idea of {\em query federation} is to provide integrated access to distributed sources on the Web.
DARQ \cite{quilitz2008querying} and SemWIQ \cite{langegger2008semantic} are two
of the first systems to support SPARQL query federation to multiple SPARQL endpoints.
They provide access to distributed RDF data sources using a mediator
service that transparently distributes the execution of queries to multiple endpoints.
Given the need to address query federation, in 2013 the SPARQL W3C working
group proposed a query federation extension for SPARQL 1.1 \cite{sparqlFedQuery}.
Buil-Aranda et al. \cite{buil2013federating} describe the syntax of this extension, formalize its semantics,
and implement a static optimization for queries that contain the OPTIONAL operator,
the most costly operator in SPARQL.

There is also a plethora of query federation engines to support efficient SPARQL query processing to
multiple endpoints.
The work in \cite{saleem2016fine} provides a comprehensive analysis, comparison,
and evaluation of a large number of SPARQL endpoint federation systems.

The ANAPSID system \cite{acosta2011anapsid} adapts query execution schedulers to data availability
and run-time conditions. It stores information about the
available endpoints and the ontologies used to describe the data in order to decompose queries
into sub-queries that can be executed by the selected endpoints, while adaptive physical
operators are executed to produce answers as soon as responses from the available
remote sources are received.
The {\em query optimizer} component of ANAPSID exploits statistics about
the distribution of values in the different datasets in order to identify the
best combination of sub-queries.

The work in \cite{montoya2012heuristic} proposes a heuristic-based approach
for endpoint federation.
Basic graph patterns are decomposed into sub-queries that can be executed by the available endpoints,
while the endpoints are described in terms of the list of predicates they contain.
Similar to ANAPSID, sub-queries are combined in a bushy tree execution plan,
while the SPARQL 1.1 federation extension is used to specify the URL of the endpoint
where the sub-query will be executed.

SPLENDID \cite{gorlitz2011splendid} is another endpoint federation system
which relies on statistical data obtained from VoID descriptions \cite{alexander2009describing}.
For triple patterns with bound variables not covered in the VoID statistics,
SPLENDID sends ASK queries to all the pre-selected data sources and removes those which fail the test.
Bind and hash joins are used to integrate the results of the sub-queries, while a dynamic programming
strategy is exploited to optimize the join order of SPARQL basic graph patterns.

ADERIS \cite{lynden2011aderis} is a query processing system for efficiently
joining data from multiple distributed endpoints.
ADERIS decomposes federated SPARQL queries into multiple source queries and
integrates the results through an adaptive join
reordering method for which a cost model is defined.

The FedX framework \cite{schwarte2011fedx} provides join processing and grouping techniques to minimize the number of requests to remote endpoints.
Source selection is performed without the need of preprocessed metadata.
It relies on SPARQL ASK queries and a cache which stores the most recent ASK requests.
The input query is forwarded to all of the data sources and those sources which pass the SPARQL ASK test are selected.
FedX uses a rule-based join optimizer which considers the number of bound variables. One of the methods we examine in this paper (UVC) is also based on the same heuristic.

Regarding more recent works, SemaGrow \cite{charalambidis2015semagrow} is a federated SPARQL querying system that uses metadata about the federated data sources to optimize query execution.
The system balances between a query optimizer that introduces little overhead, has appropriate fall backs in the absence of metadata, but at the same time produces optimal plans in many situations.
It also  exploits non-blocking and asynchronous stream processing to achieve efficiency and robustness.

Finally, Odyssey \cite{montoya2017odyssey} is a cost-based query optimization approach for endpoint federation. It defines statistics for representing both entities and links among datasets, and uses the computed statistics
to estimate the size of intermediate results.
It also exploits dynamic programming to produce an efficient query execution plan with a low number of intermediate results.

\vspace{2mm} \noindent
{\bf Our approach.}
In this work, we focus on optimizing SPARQL 1.1 and SPARQL-LD queries
through plain {\em query reordering}. The input is a query containing
two or more \service\ patterns,
and the output is a near-optimal (in terms of query execution time)
reordering of the contained \service s, i.e., an optimized {\em reordered} query.
Given the dynamic nature of Linked Data as well as
the advanced query capabilities offered by SPARQL-LD
(enabling to query any remote HTTP resource containing or returning RDF data),
we aim at providing a general query reordering method that does not
require statistics or metadata from the remote resources and that,
contrary to the aforementioned works,
can be directly applied on top of existing SPARQL 1.1 and SPARQL-LD implementation.

\subsection*{Selectivity-based Query Optimization}
Another line of research has investigated optimization methods for non-fede\-ra\-ted SPARQL queries based on selectivity estimation.

The work in \cite{stocker2008sparql} defines and analyzes heuristics for selectivity-based basic graph pattern optimization.
The heuristics range from
simple triple pattern variable counting to
more sophisticated selectivity estimation techniques that consider pre-computed triple pattern statistics.
Likewise, \cite{tsialiamanis2012heuristics} describes a set of heuristics for deciding which triple patterns
of a SPARQL query are more selective and thus it is in the
benefit of the planner to evaluate them first.
The planner tries to maximize the number of merge joins and reduce intermediate results by choosing triples patterns most likely to have high selectivity.
\cite{song2015extended} extends these works
by considering more SPARQL expressions,
in particular the operators {\tt FILTER} and {\tt GRAPH}.

In \cite{huang2011estimating} the authors study the star and chain patterns with correlated
properties and propose two methods for estimating their selectivity based on precomputed statistics.
For star query patterns, Bayesian networks are constructed to compactly
represent the joint probability distribution over values
of correlated properties, while for chain query patterns the chain histogram
is built which can obtain
a good balance between the estimation accuracy and space cost.

\vspace{2mm} \noindent
{\bf Our approach.} Similar to \cite{stocker2008sparql}, \cite{tsialiamanis2012heuristics} and \cite{song2015extended},
we exploit {\em heuristics} for selectivity estimation.
However, we focus on reordering a set of \service\ graph patterns
in order to optimize the execution of SPARQL 1.1 and SPARQL-LD queries.
Some of the heuristics we examine in this paper
are based on the results of these previous works.

\section{Query Reordering}
\label{sec:optim}

We first model query reordering as a {\em cost minimization} problem (Section \ref{subsec:modeling}).
Then we describe four heuristics-based methods for
computing the cost of a \service\ graph pattern (Section \ref{subsec:methods}).
We also discuss how we handle some special query cases (Section \ref{subsec:special}).
At the end we motivate the need for a greedy algorithm for computing a near-optimal reordering
for cases of large number of \service\ graph patterns (Section \ref{subsec:greedy}).

\subsection{Problem Modeling}
\label{subsec:modeling}
Let $Q$ be a SPARQL query and
let $S = (s_1, s_2, \dots, s_n)$ be a {\em sequence} of $n$ \service\ patterns contained in $Q$.
For a service pattern $s_i$, let $g_i$ be its nested graph pattern and
$B_i$ be the list of bindings of $Q$ {\em before} the execution of $s_i$.
Our objective is to compute a reordering $S'$ of $S$ that minimizes its {\em execution cost}.
Formally:

\begin{equation}
R^*=\argmin_{\substack{S'}} cost(S')
\end{equation}

In our case, the {\em execution cost} of a sequence of \service\ patterns $S'$
corresponds to its total execution time.
However, the execution time of a \service\ pattern $s_i \in S'$ highly depends on the query patterns that precede $s_i$,
while the bindings produced by $s_i$ affect the execution time of the succeeding \service\ patterns.
Considering the above, we can estimate $cost(S')$ as the
weighted sum of the cost of each service pattern $s_i \in S'$ given $B_i$.
Formally:
\begin{equation}
cost(S')=\sum_{i=1}^{n}{(cost(s_i | B_i) \cdot w_i)}
\end{equation}
where $cost(s_i | B_i)$ expresses the cost of \service\ pattern $s_i$ given $B_i$
(i.e., given the already-bound variables before executing $s_i$), and
$w_i$ is the weight of \service\ pattern $s_i$
which expresses the degree up to which it influences the execution time of the sequence $S'$.
We define  $w_i= \frac{n-i+1}{n}$. In this case,
for a sequence of four \service\ patterns $S' = (s_1, s_2, s_3, s_4)$,
the weights are:
$w_1 = 1.0$ (since $s_1$ influences the execution time of 3 \service\ patterns),
$w_2 = 0.75$ ($s_2$ affects 2 \service\ patterns), $w_3 = 0.5$ ($s_3$ affects 1 \service\ pattern), and $w_4 = 0.25$ ($s_4$ does not affect any other \service\ pattern).

Now, the cost of each \service\ pattern $s_i$ can be estimated
based on the {\em selectivity/restrictiveness} of its graph pattern $g_i$ given $B_i$.
Formally:
\begin{equation}
\label{eq:costFormula}
cost(s_i | B_i) = unrestrictiveness(g_i | B_i)
\end{equation}

A \service\ graph pattern that is very unrestrictive will return a large number of intermediate results (large number of bindings),
which in turn will increase the number of calls to succeeding \service\ patterns,
resulting in higher total execution time.
In the query of Listing \ref{fig:introExample} for example, a large number of bindings of the variables in the first \service\ pattern will result in many calls of the second \service.
Thus, our objective is to first execute the more restrictive \service\ patterns
that will probably return small result sets.

As proposed in \cite{stocker2008sparql} and \cite{tsialiamanis2012heuristics} (for the case of triple patterns),
the restrictiveness of a graph pattern
can be determined by the {\em number} and {\em type}
of new (unbound) variables in the graph pattern.
The most restrictive graph pattern can be considered the one containing
the less unbound variables (since fewer bindings are expected).
Regarding the type of the unbound variables, subjects can be considered more restrictive than
objects, and objects more restrictive than predicates
(usually there are more triples matching a predicate than a subject or an
object, and more triples matching an object than a subject) \cite{stocker2008sparql}.
Moreover, the number and type of joins can also affect the restrictiveness of a graph pattern since, for example, an unusual subject-predicate join will probably return less bindings.
Finally, literals and filter operators usually restrict the number of bindings and thus
increase the restrictiveness of a graph pattern.
Below, we define formulas for {\em unrestrictiveness} that consider the above factors.

\subsection{Methods for Estimating Unrestrictiveness}
\label{subsec:methods}

We examine four methods for computing the {\em unrestrictiveness cost} (Equation \ref{eq:costFormula})
of a \service\ graph pattern:
\vspace{1.0mm}
\begin{compactitem}
\item   I. Variable Count (VC)
\item   II. Unbound Variable Count (UVC)
\item   III. Weighted Unbound Variable Count (WUVC)
\item   IV. Joins-aware Weighted Unbound Variable Count (JWUVC)
\end{compactitem}

\vspace{2mm} \noindent
{\bf I. Variable Count (VC).}
The first unrestrictiveness measure simply considers the number of graph pattern variables
without considering whether they are bound or not.
For a given graph pattern $g_i$,
let $V(g_i)$ be the set of variables of $g_i$.
The unrestrictiveness of $g_i$ can be now defined as:
\begin{equation}
unrestrictiveness(g_i | B_i) = |V(g_i)|
\end{equation}
With the above formula, more variables in a graph pattern means higher unrestrictiveness score.
Consider for example the query in Listing \ref{fig:vc_example}.
The second \service\ pattern contains one variable and is more likely to retrieve a
smaller number of results than the first one which contains three variables.
Thus the second \service\ pattern is more restrictive and should be executed first.

\begin{figure}[th]
\vspace{-4mm}
\centering \scriptsize
\begin{Verbatim}[frame=lines,numbers=left,numbersep=1pt]
 SELECT * WHERE {
   SERVICE <http://resource1> { ?s ?p ?o }
   SERVICE <http://resource2> { ?s a :fish } }
\end{Verbatim}
\vspace{-4mm}
\caption{Example SPARQL query for VC reordering.}
\label{fig:vc_example}
\vspace{-4mm}
\end{figure}

\vspace{2mm} \noindent
{\bf II. Unbound Variable Count (UVC).}
A \service\ pattern containing many new unbound variables is more likely to retrieve a higher
number of results compared to a \service\ pattern with less unbound variables.
Thereby, we can also consider the set of binding $B_i$ before the execution of a \service\ pattern $s_i$.
Let first $V^u(g_i, B_i)$ be the set of new (unbound) variables of $g_i$ given $B_i$.
The unrestrictiveness of $g_i$ can be now defined as:
\begin{equation}
unrestrictiveness(g_i | B_i) = |V^u(g_i, B_i)|
\end{equation}
Listing \ref{fig:uvc_example} shows an example for this case.
After the execution of the first \service\ pattern, we should better run the third one
since all its variables are already bound. The second \service\ pattern contains one unbound variable,
although its total number of variables is less than those of the third \service\ pattern.

\begin{figure}[th]
\vspace{-3mm}
\centering \scriptsize
\begin{Verbatim}[frame=lines,numbers=left,numbersep=1pt]
  SELECT * WHERE {
    SERVICE <http://resource1> {
      <http://entity1> :birthPlace ?place1 ; :friend ?entity2 ; :workPlace ?place2 }
    SERVICE <http://resource2> { ?entity2 a ?type } 
    SERVICE <http://resource3> { ?entity2 :birthPlace ?place1 ; :workPlace  ?place2 } }
\end{Verbatim}
\vspace{-4mm}
\caption{Example SPARQL query for UVC reordering.}
\label{fig:uvc_example}
\vspace{-4mm}
\end{figure}

\vspace{2mm} \noindent
{\bf III. Weighted Unbound Variable Count (WUVC).}
The above formulas do not consider the type of the unbound variables in the graph pattern,
i.e., whether they are in the subject, predicate or object position in the triple pattern.
For a graph pattern $g_i$ and a set of bindings $B_i$,
let $V^u_s(g_i, B_i)$, $V^u_p(g_i, B_i)$ and $V^u_o(g_i, B_i)$
be the set of subject, predicate and object unbound variables in $g_i$, respectively.
Let also $w_s$, $w_p$ and $w_o$ be the weights for subject, predicate and object variables, respectively.
The unrestrictiveness of $g_i$ can be now defined as:
\begin{equation}
\small
unrestrictiveness(g_i | B_i) = |V^u_s(g_i, B_i)| \cdot w_s + |V^u_p(g_i, B_i)| \cdot w_p + |V^u_o(g_i, B_i)| \cdot  w_o
\end{equation}

According to \cite{stocker2008sparql}, subjects are in general more restrictive than objects and objects are more restrictive than predicates,
i.e., there are usually more triples matching a predicate than an object,
and more triples matching an object than a subject.
When considering variables, selectivity is opposite:
a subject variable may return more bindings than an object variable
and an object variable more bindings than a predicate variable.
Consider for example the query in Listing \ref{fig:wuvc_example}.
The subjects having {\em Greece} as the birth place (1st \service\ pattern) are expected to be more than the friends of {\em George} (2nd \service\ pattern), while the friends of {\em George}
are expected to be more than the different properties that connect {\em George} with {\em Nick} (3rd \service\ pattern).
Thus, one can define weights so that $w_s > w_o > w_p$.
Based on the distribution of subjects, predicates and objects in
a large Linked Data dataset of more than 28 billion triples (gathered from more than 650 thousand sources) \cite{fernandez2017lod},
we define the following weights: $w_s = 1.0$,  $w_o = 0.8$, $w_p = 0.1$.
Moreover, if a variable exists in more than one triple pattern position
(e.g., both as subject or object), we consider it as being in the more restrictive position.

\begin{figure}[th]
\vspace{-5mm}
\centering \scriptsize
\begin{Verbatim}[frame=lines,numbers=left,numbersep=1pt]
 SELECT * WHERE {
   SERVICE <http://resource1> { ?entity1 :birthPlace :Greece }
   SERVICE <http://resource2> { <http://George> :friend ?entity1 }
   SERVICE <http://resource3> { <http://George> ?p <http://Nick> } }
\end{Verbatim}
\vspace{-4mm}
\caption{Example SPARQL query for WUVC reordering.}
\label{fig:wuvc_example}
\vspace{-4mm}
\end{figure}

\vspace{2mm} \noindent
{\bf IV. Joins-aware Weighted Unbound Variable Count (JWUVC).}
When a graph pattern contains joins, its restrictiveness is usually increased
depending on the number and type of joins (star, chain, or unusual join) \cite{huang2011estimating}.
For a graph pattern $g_i$,
let $J_*(g_i)$, $J_{\rightarrow}(g_i)$, and $J_{\times}(g_i)$
be the number of star, chain, and unusual joins in $g_i$, respectively.
We consider the subject-subject and object-object joins as {\em star joins},
the object-subject and subject-object as {\em chain joins},
and all the others as {\em unusual joins}.
Let also $j_*$, $j_{\rightarrow}$ and $j_{\times}$ be the weights for
star, chain, and unusual joins, respectively.
Based on the assumption that, in general, unusual joins are much more restrictive than chain joins,
and chain joins are more
restrictive than star joins \cite{tsialiamanis2012heuristics},
one can define weights so that $j_{\times} > j_{\rightarrow} > j_*$.
We define: $j_{\times} = 1.0$,  $j_{\rightarrow} = 0.6$, $j_* = 0.5$.
The following unrestrictiveness formula considers both
the number and the type of joins in the graph pattern $g_i$:

\begin{equation}
\small
unrestrictiveness(g_i | B_i) = \frac{|V^u_s(g_i, B_i)| \cdot w_s + |V^u_p(g_i, B_i)| \cdot w_p + |V^u_o(g_i, B_i)| \cdot  w_o}
{1 + J_*(g_i) \cdot j_* + J_{\rightarrow}(g_i) \cdot j_{\rightarrow} + J_{\times}(g_i) \cdot j_{\times}}
\end{equation}

Listing \ref{fig:jwuvc_example} shows an example for this case.
The first \service\ pattern contains a star join,
the second a chain join, and the third an unusual join.
The unusual join will probably return fewer results than the star and chain joins.

\begin{figure}[th]
\vspace{-2mm}
\centering \scriptsize
\begin{Verbatim}[frame=lines,numbers=left,numbersep=1pt]
 SELECT * WHERE {
   SERVICE <http://resource1> { ?ent1 :birthPlace :Greece ; :workPlace :Germany }
   SERVICE <http://resource2> { <http://George> :friend ?ent1 . ?ent1 :friend <http://Nick> }
   SERVICE <http://resource3> { <http://George> ?p <http://Nick> . ?p :label "best friend" } }
\end{Verbatim}
\vspace{-4mm}
\caption{Example SPARQL query for JWUVC reordering.}
\label{fig:jwuvc_example}
\vspace{-4mm}
\end{figure}

In Section \ref{sec:eval} we evaluate the effectiveness of the above four methods
on finding the optimal, in terms of query execution time, query reordering.

\subsection{Handling of Special Cases}
\label{subsec:special}

{\em Query plans with same cost.} In case the lowest unrestrictiveness cost is the same
for two or more query reorderings, we consider the number of {\em literals} and
{\em filter} operators contained in the graph patterns.
Literals are generally considered more selective than URIs \cite{tsialiamanis2012heuristics},
while a filter operator limits the bindings of the filtered variable and thus increases
the selectivity of the corresponding graph pattern \cite{stocker2008sparql}.
Thus, we count the total number of literals and filter operators in each \service\ pattern, and consider it
when we get query plans with the same unrestrictiveness cost.
If the corresponding \service\ patterns contain the same number of literals and filter operators,
then we maintain their original ordering, i.e., we order them based on their order in the input query.

\vspace{1mm} \noindent
{\em SERVICE within OPTIONAL.} In case a \service\ call is within an {\em optional} pattern,
then we separately reorder the \service\ patterns that exist before and after it.
An optional pattern requires a left outer join and thus changing its order
can distort the query result.

\vspace{1mm} \noindent
{\em Variable in SERVICE clause.} If a \service\ clause contains a
variable instead of a URI, we should ensure that this variable gets bound before
the execution of the \service\ pattern.
Thereby, during reordering we ensure that all other \service s containing this variable in their graph patterns
are placed before the \service\ pattern having the variable in its clause.

\vspace{1mm} \noindent
{\em Projection variables.}
The set of variables that appear in the {\tt SELECT} clause of a \service\ pattern
are called the projection variables.
Since these are part of the answer and affect the size of the bindings,
we only consider these variables in all the proposed formulas.

\vspace{1mm} \noindent
{\em UNION operator, nested patterns, combination of triple and SERVICE patterns.}
In this work we do not study the case of queries containing the {\tt UNION} operator
as well as nested patterns.
Such queries require the reordering of groups of \service\ patterns which
is not currently supported by our implementation.
In addition, our implementation does not yet support the case of queries containing both
triple patterns (that query the \lq\lq{}{}local\rq\rq{}{} endpoint) and \service\ patterns.
We leave the handling of these cases as part of our future work.

\subsection{Computing a near-optimal query-execution plan}
\label{subsec:greedy}

Computing the unrestrictiveness score for all the different query reorderings
may be prohibitive for large number of \service\ patterns,
since the complexity is $n!$ (where $n$ is the query's number of \service\ patterns).
This applies in all the proposed optimization methods apart from VC where no all permutations are needed to be computed.
For example, for queries with 5 \service\ patterns there are $5!$ (=720) different permutations,
however for 10 \service\ patterns this number is increased to more than 3.6 million permutations
and for 15 to around 1.3 trillion.

Table \ref{tbl:timeComputePlan} shows the time required for computing the reordering with the lowest cost for different number of \service\ patterns using the JWUVC method (the time is almost the same for also UVC and WUVC).
Our implementation (cf. Footnote \ref{footnote:sourceCode}) is in Java and uses Apache Jena for decomposing the SPARQL query, while we run the experiments in an ordinary computer
with processor Intel Core i5 @ 3.2Ghz CPU, 8GB RAM and running Windows 10 (64 bit).

\begin{table}
\vspace{-5mm}
\centering
  \caption{Time to compute the reordering with the lowest cost for different number of \service\ patterns in a SPARQL query.}
  \vspace{1.5mm}
  \label{tbl:timeComputePlan}
  \begin{tabular}{cc}
    \toprule
    Number of \service\ patterns ~&~ Time\\
    \midrule
    5       ~&~ 8 ms    \\
    6       ~&~ 22 ms   \\
    7       ~&~ 89 ms   \\
    8       ~&~ 290 ms  \\
    9       ~&~ 2.4 sec     \\
    10      ~&~ 25 sec   \\
    11      ~&~ 6 min     \\
    12      ~&~ 67 min     \\
    13      ~&~ $>$5 hours     \\
    14      ~&~ $>$5 hours     \\
  \bottomrule
\end{tabular}
\vspace{-3.5mm}
\end{table}

We see that the time is very high for queries with many \service\ patterns. For example, more than 1 hour is required for just finding the reordering with the lowest cost for a query with 12 \service\ patterns. This illustrates the need for a cost-effective approach
which can find a near-optimal query execution plan without needing to check all the different permutations.
We adopt a greedy algorithm starting with the \service\ pattern
with the smaller unrestrictiveness score (local optimal choice) and
continuing with the next \service\ pattern with the smaller score, considering
at each stage the already bound variables of the previous stages.
To find the local optimal choice, we can use any of the proposed unrestrictiveness formulas.
Considering the UVC formula for example, in the query of Listing \ref{fig:uvc_example2}
the greedy algorithm first selects the 2nd \service\ pattern since it contains
only 1 variable. In the next stage, it selects the 3rd \service\ pattern which
contains 2 unbound variables, fewer than those of the 1st \service\ pattern.

\begin{figure}[th]
\vspace{-4mm}
\centering \scriptsize
\begin{Verbatim}[frame=lines,numbers=left,numbersep=1pt]
 SELECT * WHERE {
  SERVICE <http://resource1> { ?ent1 :birthPlace ?place1 ; :workPlace ?place2 ; :friend ?ent2 }
  SERVICE <http://resource2> { ?ent2 a :Actor }
  SERVICE <http://resource3> { ?ent2 :birthPlace ?place1 ; :workPlace  ?place2 } }
\end{Verbatim}
\vspace{-4mm}
\caption{Example SPARQL query for choosing a near-optimal query plan.}
\label{fig:uvc_example2}
\vspace{-4mm}
\end{figure}

\section{Evaluation}
\label{sec:eval}

We evaluated the effectiveness of the proposed query reordering methods using
real federated queries from the {\em LargeRDFBench} \cite{saleem2018largerdfbench} dataset\footnote{\url{https://github.com/dice-group/LargeRDFBench}}.
From the provided 32 SPARQL 1.1 queries, we did not consider 10 queries
that make use of the {\tt UNION} operator
(it is not currently supported by our implementation) and 5 \q{large data} queries
(due to high memory requirements).
To consider larger number of possible query permutations,
and since some of the queries contain only 2 \service\ patterns,
we removed the {\tt OPTIONAL} operators keeping though the embedded \service\ pattern(s).\footnote{Although
this transformation changes the query results, it does not affect the objective of our evaluation.}
For instance, we transformed the query:\\
\small
\indent {\tt SELECT * WHERE \{ SERVICE <ex1> \{..\} OPTIONAL \{ SERVICE <ex2> \{..\} \} \}} \\
\normalsize
to the query:\\
\small
\indent {\tt SELECT * WHERE \{ SERVICE <ex1> \{..\} SERVICE <ex2> \{..\} \}} \\
\normalsize
\vspace{-2mm}

The final evaluation dataset contains 17 queries of varying complexity (each one containing at least two \service\ patterns), while their \service\ patterns require access to totally 7 remote SPARQL endpoints.
Note that there is no benchmark for SPARQL-LD, however this does not affect the objective of our evaluation since the proposed methods do not distinguish between SPARQL 1.1 and SPARQL-LD queries (a SPARQL endpoint can be considered an HTTP resource containing all the endpoint's triples).

For each query, we found the optimal reordering by computing the execution
time of all possible permutations (average of 5 runs).
Then, we examined the effectiveness of the proposed optimization methods
(VC, UVC, WUVC, and JUWVC, as described in Section \ref{subsec:methods})
on finding the optimal query execution plan.
Figure \ref{fig:evalResults} shows the results.
VC finds the optimal query plan in 8/17 queries (47\%),
UVC in 10/17 queries (59\%), WUVC in 9/17 queries (53\%), and JUWVC in 15/17 queries (88\%).
We notice that the JUWVC method, which considers the number and type of joins,
achieves a very good performance.
Given the infrastructure used to host the SPARQL endpoints in our
experiments\footnote{2x Intel Xeon CPU E5-2630 @ 2.30GHz, 6-core, 384GB RAM.},
query reordering using JUWVC achieves a very large decrease of the query execution time for many of the queries
(for example, from minutes to some seconds for the queries S4, S10, S12, C7, C10).

\renewcommand{\figurename}{Figure}
\setcounter{figure}{0}
\begin{figure}[h]
	\vspace{-2mm}
	\includegraphics[scale=.41]{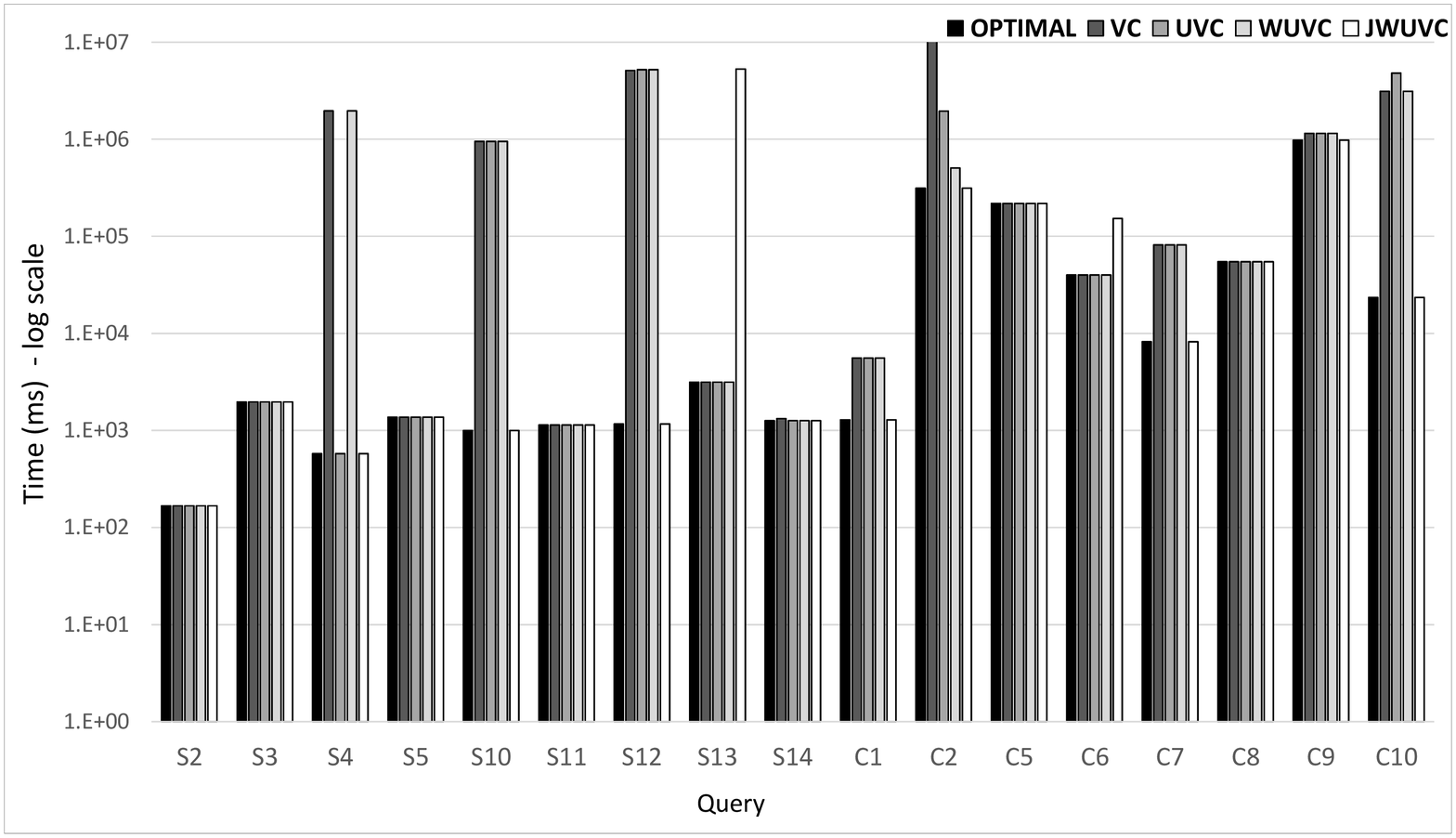}
	\vspace{-4mm}
	\caption{Effectiveness of the different query reordering methods
		(VC: variable count; UVC: unbound variable count; WUVC: weighted unbound variable count; JWUVC: joins-aware weighted unbound variable count).}
	\label{fig:evalResults}
	\vspace{-5mm}
\end{figure}

JUWVC fails to find the optimal query plan for the queries S13 and C6,
which both contain 2 \service\ patterns. The first \service\ pattern of S13
contains 1 star join and the second 2 star joins.
As regards C6, its first \service\ pattern contains 1 star join and 1 chain join,
and its second 5 star joins.
In both queries, although the second \service\ pattern contains more joins than the first \service\ pattern,
it returns larger number of bindings and this increases the number of calls to the first remote endpoint
and thus the overall query execution time.
Note that, without exploiting dataset statistics,
such cases are very difficult to be caught by an unrestrictiveness formula.

As regards the effectiveness of the greedy algorithm which avoids computing the cost of all possible permutations (cf. Section \ref{subsec:greedy}),
it manages to find the reordering with the lowest cost using JUWVC in 16/17 queries (94\%).
It fails for the query C2, however the returned reordering is very close to the optimal (the difference is only a few milliseconds).

One of the limitations of such a fully heuristics-based method is that it is practically impossible to always find the optimal query plan. However, this is the case also for methods that pre-compute and exploit metadata and statistics from the remote resources, or which make use of caching. The reason is that the Web of Data is a huge and constantly evolving information space, meaning that we may always need to query a new, unknown resource discovered during query execution. 
A solution to this problem is the exploitation of VoID \cite{alexander2009describing}, in particular the publishing of a rich VoID file alongside each resource. In this case, an optimizer can access (and exploit for query reordering) such VoID descriptions at query execution time, considering though that all publishers follow a common pattern for publishing these VoID files.\footnote{\url{https://www.w3.org/TR/void/#void-file}}

\section{Conclusion}
\label{sec:concl}

We have proposed and evaluated a set of fully heuristics-based query reordering methods
for federating queries in SPARQL 1.1 and SPARQL-LD.
The proposed methods reorder a set of \service\ graph patterns based on their selectivity (restrictiveness)
and do not require the gathering and use
of statistics or metadata from the remote resources.
Such an approach is widely-applicable and can be exploited on top of existing SPARQL 1.1
and SPARQL-LD implementations.

Since the new query functionality offered by SPARQL-LD
(allowing to query any HTTP resource containing RDF data)
can lead to queries with large number of \service\ patterns which in turn can dramatically increase the time to find the optimal reordering,
we proposed the use of a simple greedy algorithm
for finding a near-optimal query execution plan without checking all possible query reorderings.
The results of an experimental evaluation using an existing benchmark showed that a query reordering method which considers the number and type of unbound variables and the number and type of joins achieves the optimal query plan in 88\% of the examined queries,
resulting in a large decrease of the overall query execution time (from minutes to a few seconds in many cases).
Regarding the greedy algorithm, it has an accuracy of 94\% in finding the reordering with the lowest cost.

As part of our future work, we plan to offer a holistic query reordering approach which will cover any type of federated queries.
This involves the handling of queries containing {\tt UNION} and nested graph patterns,
as well as queries which combine triple and \service\ patterns.
We also plan to offer this query reordering functionality as a web service, allowing for on-the-fly query optimization.

\subsection*{Acknowledgements}
The work was partially funded by the European Commission for the ERC Advanced Grant ALEXANDRIA under grant No. 339233.

\bibliographystyle{splncs04}
\bibliography{sparql-ld_optim}

\end{document}